\documentstyle[prl,twocolumn,aps,epsf,epsfig,amsmath,amssymb]{revtex}
\begin{document}

\twocolumn[
\hsize\textwidth\columnwidth\hsize\csname@twocolumnfalse\endcsname
\draft

\title{Metallic behavior in Si/SiGe 2D electron systems}
\author{E. H. Hwang and S. Das Sarma}
\address{Condensed Matter Theory Center, 
Department of Physics, University of Maryland, College Park,
Maryland  20742-4111 } 
\date{\today}
\maketitle

\begin{abstract}
We calculate the temperature, density, and parallel magnetic field
dependence of low temperature electronic resistivity in 2D
high-mobility Si/SiGe quantum structures, assuming the
conductivity limiting mechanism to be carrier scattering by screened
random charged Coulombic impurity centers. We obtain comprehensive
agreement with existing
experimental transport data, compellingly establishing that the
observed 2D metallic behavior in low-density Si/SiGe systems arises
from the peculiar nature of 2D screening of long-range impurity
disorder. In particular, our theory correctly predicts 
the experimentally observed metallic
temperature dependence of 2D
resistivity in the fully spin-polarized system.

\noindent
PACS Number : 71.30.+h; 73.40.Qv
\end{abstract}
\vspace{0.5cm}
]

\newpage

The two-dimensional (2D) ``metallicity'' is by now a well-established
ubiquitous low-temperature phenomena \cite{one} where a relatively
high-quality (i.e. low disorder) 2D carrier (either electron or hole)
system confined in semiconductor heterostructures exhibits a pronounced
temperature ($T$) and/or parallel (i.e. in the 2D plane) magnetic field
($B_{\|}$) dependent resistivity $\rho(T;B_{\|})$ in the dilute or low
carrier density regime. The strength of the metallicity, i.e. the
relative magnitude of the variation of $\rho(T)$ with $T$,
is found to depend on the specific 2D material involved as
also does the precise low-density regime where the strong metallicity
manifests itself. For example, in 2D Si MOSFETs the metallicity occurs
\cite{two} around $10^{11}cm^{-2}$ with $\rho(T)$ changing by as much
as a factor of $3-4$ as temperature changes from 50 mK to 5K (at
$B=0$) whereas in 2D n-GaAs electron systems the metallicity is
observed \cite{three} for density well below $10^{10}cm^{-2}$ with
$\rho(T)$ changing only by about 25\% as $T$ changes from 50 mK to a
few K. The 2D metallicity has been experimentally observed in
essentially all existing semiconductor-based 2D systems, most notable
(and most extensively studied) being electrons in Si MOSFETs, holes in
p-GaAs 2D structures \cite{four} and very recently, electrons in
n-GaAs 2D structures \cite{three}.

One of the more important systems to manifest 2D
metallic behavior is the 2D electron system in modulation doped
Si/SiGe structures \cite{five,six,seven,eight} where the electrons are
confined by the Si-SiGe interface potential barrier on the
Si side of the heterostructure and the modulation doping delta layer
(contributing the 2D electrons) is typically placed 100 -- 200 \AA \;
inside the insulating SiGe side. The relatively impurity-free high
quality of the materials (both Si and SiGe), and the lack of interface
impurities and roughness (at the Si-SiGe interface) make the Si/SiGe
2D electron system essentially ideal from the perspective of
understanding 2D metallic behavior. In particular, the dominant
low-temperature (below 5K or so where phonon scattering should be
exponentially suppressed in silicon) scattering mechanism in Si/SiGe
2D electron system is thought to be scattering by the remote dopants
in the modulation delta layer which, being far (i.e. 100 -- 200 \AA)
away from the 2D electrons residing inside Si, leads to very high, by
more than an order of magnitude higher than the corresponding
Si MOSFETs, carrier mobilities. The high quality of the system 
and the simple isotropic
parabolic 2D energy dispersion (in contrast to the p-GaAs based 2D
hole systems which suffer from the multiband valence band structure,
strong 2D anisotropy, strong spin-orbit coupling, inter-spin-split-band
scattering, and strong phonon scattering leading to considerable
complications in its 2D metallicity) make Si/SiGe based 2D electron
systems an excellent candidate
for studying 2D metallic behavior. In
particular, 2D electrons in Si/SiGe structures combine, from the 2D
metallic physics perspective, the advantages 
of  n-Si MOSFETs (i.e. strong metallicity)
with the advantages of the n-GaAs system
(i.e., high mobility and low density)
without the n-GaAs disadvantage of weak  metallicity 
and
phonon scattering problems. Metallicity has been observed \cite{eight}
in 2D Si/SiGe systems down to densities as low as $3 \times 10^{10}
cm^{-2}$, and the relevant interaction parameter ($r_s$) is typically
5 -- 10 for this system.

In the context of 2D metallic physics it is therefore of considerable
interest that recently two independent groups have reported detailed
experimental data on the temperature, density, and parallel magnetic
field dependence of 2D resistivity in Si/SiGe confined electron
systems. The reported Si/SiGe experimental results, particularly in
the presence of a parallel magnetic field, have significant {\it
  qualitative} difference with the corresponding experimental results
in n-Si MOSFETs, warranting a theoretical
investigation. We provide, in this Communication, such a transport
theory for 2D electrons in Si/SiGe structures, finding good agreement
between our theoretical and the existing experimental results.

We use a minimal zeroth-order theoretical model \cite{nine} to
describe the 2D transport properties, keeping the total number of
unknown parameters to a minimum. We neglect all phonon scattering
effects although it is fairly straightforward to include them, mainly
because 
our theoretical estimate shows phonon scattering to be
negligible for 2D electrons in Si/SiGe structures in the $T<5K$ regime
of interest to us. We assume that the 2D carrier conductivity is
entirely limited by screened impurity scattering, where the disorder
arises from random background charged impurity centers
(i.e. unintentional dopants) and the ionized dopants (randomly
distributed) in the modulation dopant layer. The background
unintentional impurity density is extremely low ($\sim 10^{14}
cm^{-3}$, which is equivalent to a 2D density of $10^8 cm^{-2}$), as 
implied by the high achieved mobilities ($\sim 7\times 10^5 cm^2/Vs$)
in gated Si/SiGe quantum well structures. The 
2D resistivity is given in the Boltzmann theory by 
$\rho = {m}/{ne^2 \langle \tau \rangle}$,
where $n$, $m$, $\tau$ are respectively
the 2D carrier density, the carrier effective mass, the transport
relaxation time, and
\begin{equation}
\langle \tau \rangle = {\int d\epsilon \epsilon
  \tau(\epsilon) \left (-\frac{\partial f(\epsilon)}{\partial 
  \epsilon} \right )} / {\int d\epsilon \epsilon
   \left (-\frac{\partial f(\epsilon)}{\partial
  \epsilon} \right )}, 
\end{equation}
with $f(\epsilon)$ being
the Fermi distribution function and
\begin{eqnarray}
\frac{1}{\tau(\epsilon_{\bf k})} = \frac{2\pi}{\hbar} \int dz \int
\frac{d^2 k'}{(2\pi)^2} & &N_i(z) \left |u^{e-i}({\bf k}-{\bf k'},z)
\right |^2 \nonumber \\
&\times & (1-\cos \theta_{\bf kk'}) \delta (\epsilon_{\bf
  k}-\epsilon_{\bf k'}).
\end{eqnarray}
Eq. (1)
indicates a thermal average over the energy dependent relaxation time
$\tau(\epsilon)$, with $\epsilon = k^2/2m$ being the usual
parabolic 2D electron energy dispersion.
In Eq. (2), $u^{e-i}(q;z)$ is the (finite-temperature) screened
electron-impurity interaction for 2D momentum transfer of $q$ and
$N_i(z)$ is the random charged impurity density in the direction ($z$)
normal to the 2D ($x$-$y$) plane of confinement of the electron layer
(the $z=0$ being the Si/SiGe interface plane), which is given in our
model by
$N_i(z) = N_i^{3D} + n_i \delta(z-z_i)$,
where $N_i^{3D}$ is the background 3D charged impurity density and
$n_i$ is the 2D charged dopant density in the modulation doping
$\delta$-layer at $z=z_i$. We take $z_i$ to coincide with the
modulation doping layer in the experimental Si/SiGe samples. The key
quantity in Eqs. (1)--(2) is the screened charged impurity disorder
potential $u^{e-i}$, which we take to be
\begin{equation}
u^{e-i}(q,z) = v^{e-i}(q;z)/\varepsilon(q),
\end{equation}
where $v^{e-i}$ is the bare Coulomb potential for electron-charged
impurity interaction, and $\varepsilon(q) = 1-v(q)\Pi(q)$ is the RPA
dielectric screening function due to the 2D electrons themselves, where
$v(q)$ is the 2D bare Coulomb electron-electron interaction, and
$\Pi(q)\equiv \Pi(q;T)$ is the 2D finite-temperature and finite wave
vector 
polarizability function. All our calculations include the realistic
quasi-2D quantum form factor effects, arising from the finite width of
the 2D layer in $z$-direction, which enter the expressions for $v(q)$
and $v^{e-i}(q)$, that are of considerable quantitative importance. All
quantities entering the theory are known except for $N_i^{3D}$ and
$n_i$, which are the only two  parameters of the model. We set
$N_i=2\times 10^{14}cm^{-3}$, $n_i=3.5\times 10^{10}cm^{-2}$, $z_i=100 \AA$
throughout our calculations, instead of using them as adjustable
parameters since our interest here is a qualitative (and perhaps
semiquantitative) understanding of 2D metallicity in Si/SiGe
systems. Note that there is little scattering by interface impurities and
interface roughness in  high-mobility Si/SiGe 2D systems,
which is their distinguishing feature compared with Si MOS systems,
leading to the extraordinarily (a factor of 20 -- 40 higher than in Si
MOSFETs) high 2D mobility.

\begin{figure}
\epsfysize=2.3in
\centerline{\epsffile{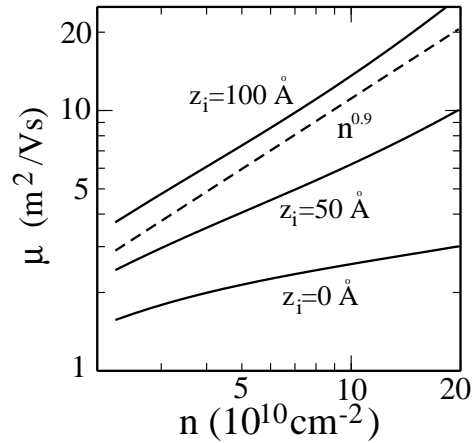}}
\vspace{0.5cm}
\caption{
The calculated 2D mobility of Si/SiGe 2D systems at 0.3K as a function of
carrier density for different  values of $z_i$. Dashed line represents
$\mu \propto n^{0.9}$.
}
\label{fig1}
\end{figure}

In Fig. 1 we show our calculated 2D mobility, $\mu = 1/ne\rho$, as a
function of 2D electron density at $T=0.3K$  for a number of different
values of $z_i$, the position of the 2D dopant layer.
The calculated mobility for $z_i=100 \AA$ agrees well (quantitatively
for $n > 0.5 \times 10^{11} cm^{-2}$) with the density-dependent
mobility measured experimentally \cite{eight} which also (in agreement
with Fig. 1) manifests a $\mu \propto n^{0.9}$ dependence on density at
higher densities in contrast to $\mu \propto n^{0.3}$ behavior observed
in Si MOSFETs (i.e., our $z_i=0$ result corresponding to scattering
by interface impurities). This establishes the validity of our model for
the experimental Si/SiGe 2D systems.

In Fig. 2 -- 4 we show our calculated results for the 2D
resistivity in Si/SiGe systems.
In Fig. 2 we show the zero-field ($B_{\|}=0$) temperature dependence
of the 2D resistivity for a number of carrier densities in the 2D
metallic phase. The theoretical results are in excellent qualitative
agreement with experimental observations \cite{five,six,seven,eight}
with the maximum temperature induced fractional change in the
resistivity, $\Delta\rho/\rho_0$ where $\rho = \rho_0 + \Delta
\rho(T)$ with $\rho_0 = \rho(T\rightarrow 0)$, being about 2 at lower
densities in the $T \approx 100 mK - 5K$ range. At higher densities
($>10^{11}cm^{-2}$), $\Delta \rho/\Delta_0$ is, consistent with the
experimental data, only about 50\% (in the $0-5$K range). 
The relatively weaker temperature dependence in Si/SiGe 2D system
compared with Si MOS system
arises entirely from the suppressed role of
$2k_F$-scattering in the modulation doped Si/SiGe systems -- in fact,
the Si/SiGe system will manifest very strong 
(stronger than Si MOSFET) 2D metallicity if its mobility would be
limited by interface scattering as in Si MOSFETs (or purely by the
background impurities as in  p-GaAs and n-GaAs 2D
systems). It is well-known that the strong metallicity in the
transport properties of low-disorder 2D systems arises from the
temperature induced suppression of the $2k_F$-Kohn anomaly in
screening (i.e. the $T=0$ cusp at $q=2k_F$ in the 2D polarizability
function), and the overall strong suppression of $2k_F$ scattering in
the modulation doped structures [due to the presence of the
$e^{-2k_F|z_i|}$ factor in the electron-impurity bare interaction
$v^{e-i}(q;z)$ in Eq. (3)] leads to relatively weaker 2D metallicity. In
fact, the temperature dependence would be substantially weaker than
that in Fig. 2 (and therefore substantially weaker than the
experimental observation) if we set the background impurity density
$N_i^{3D}$ to be zero in our calculation. The temperature dependence
(Fig. 2) of $\rho(T;n)$ is thus a simple diagnostic for the strength of
background impurity scattering in Si/SiGe system just as the density
dependence (Fig. 1) is a diagnostic for the remote impurity
scattering. We also mention the ``almost parallel'' behavior of
$\rho(T)$ at various densities in Fig. 2, which has been
experimentally seen \cite{eight}.

\begin{figure}
\epsfysize=2.3in
\centerline{\epsffile{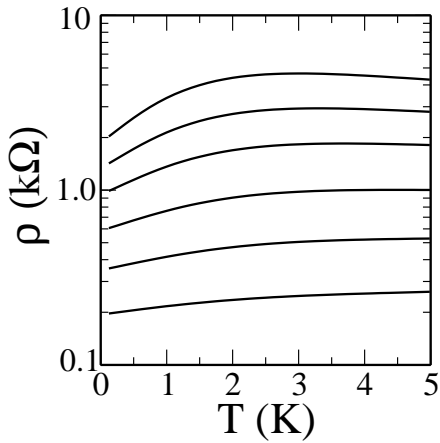}}
\vspace{0.5cm}
\caption{
Calculated resistivity as a function of temperature, $\rho(T)$, of 2D
Si/SiGe systems  
for various densities $n =$4.2, 5.15, 6.3, 8.2, 10.8 14.5$\times
10^{10}cm^{-2}$ (top to bottom). 
}
\label{fig2}
\end{figure}

In Fig. 2 $\rho(T)$ shows a nonmonotonicity (particularly prominent
at lower densities) where it increases with $T$ at first and then
decreases with increasing $T$ after going through a shallow maximum at
a characteristic density-dependent temperature $T_p(n)$ where
$d\rho/dT$ changes sign, i.e. $d\rho/dT =0$ at $T=T_p$. Our
theoretical $T_p(n)$ in Fig. 2 scales approximately with the Fermi
temperature $T_F$, but is not identical to $T_F$. This is precisely
the experimental observation \cite{eight}. This non-monotonicity,
which is obviously a non-asymptotic finite temperature phenomenon,
arises from the competition between an increasing and decreasing
contribution to $\rho(T)$ as defined by Eqs. (1) -- (3). The temperature
dependence of the dielectric screening function in Eq. (3), with
screening (particularly the strongly resistive $2k_F$-scattering)
decreasing with increasing temperature, gives rise to increasing
effective disorder with increasing $T$, and hence increasing $\rho(T)$
with $T$. At higher temperatures the thermal averaging in Eq. (1),
which always leads to $\rho(T)$ decreasing with $T$, becomes
quantitatively more important, giving rise eventually to a $\rho(T)$
decreasing with $T$ for $T>T_p$. In the classical non-degenerate
regime the thermal averaging {\it always} wins out and impurity
scattering limited $\rho(T)$ necessarily decreases with increasing $T$
(which is why we earlier called this nonmonotonicity somewhat
simplistically ``the quantum-classical crossover'', generating some
confusion on this matter), but thermal averaging {\it always}
(i.e. even in the quantum $T \ll T_F$ regime) leads to a negative
sub-leading contribution to $\rho(T)$ going as
$O(T/T_F)^2$, whereas the screening contributes the leading order term
going as $O(T/T_F)$, so in the $T/T_F \rightarrow 0$ limit the
increasing contribution arising from $2k_F$ screening wins
out. Depending on the details of the system, $T_p(n)$ could be quite
low (e.g. much lower than $T_F$ in weakly screening systems) although
it should scale approximately with $T_F$. Our theoretical $T_p(n)$ in
Fig. 2 is roughly of the order of the experimental $T_p$, but the
theoretical $T_p$ is somewhat higher quantitatively than the
experimental $T_p$. Many details (e.g. phonon scattering, exact
charged impurity distribution, the confinement potential, the
effective mass, etc.) affect the precise value of $T_p$, and
therefore a quantitative comparison between theory and experiment is
not particularly meaningful.

In Figs. 3 and 4 we show our calculated results for the 2D
magnetoresistance in Si/SiGe system in the presence of a parallel
magnetic field $B_{\|}$ which polarizes electron spins,
consequently weakening screening considerably (i.e. $B_{\|}$ acts
similar to finite $T$ effects). Of the three physical mechanisms
\cite{ten} that affect 2D parallel field magnetoresistance, we have
ignored the magneto-orbit effect \cite{eleven} because it is
negligibly small in the Si system at the applied parallel field values
of experimental interest. We include  the
other two (competing) parallel field corrections, both arising from
the electronic spin polarization induced by the applied field (and
therefore both saturating for $B_{\|} \ge B_c$ where the system is
completely spin polarized) --- these being the spin-polarization
induced weakening of screening (leading to positive magnetoresistance)
and the spin-polarization induced enhancement of the 2D Fermi wave
vector $k_F$ (leading to negative magnetoresistance). The Si 2D system
of interest to us is a strongly screening system with $q_{TF}/2k_F \gg
1$, where $q_{TF}$ is the 2D screening wave vector, and therefore the
screening effect is by far the dominant parallel field effect,
producing a strong positive magnetoresistance apparent in Figs. 3 and 4
(and in the experiments).

\begin{figure}
\epsfysize=2.3in
\centerline{\epsffile{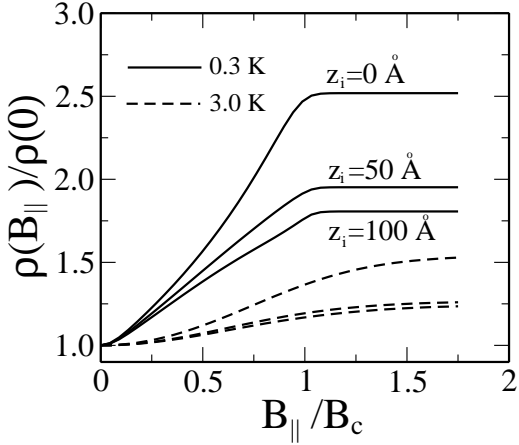}}
\vspace{0.5cm}
\caption{
Calculated magnetoresistance as a function of in-plane magnetic field
for different values of the impurity separation $z_i=0$, 50, 100 \AA \; 
at a fixed 2D
density $n=10^{11}cm^{-2}$ and at $T=0.3K$ (solid lines) and 3.0K
(dashed lines).
}
\label{fig3}
\end{figure}

In Fig. 3 we show the calculated magnetoresistance
$\rho(B_{\|})/\rho(0)$ for different values of the impurity separation
$z_i$. As expected, larger values of $z_i$ drastically reduce the
maximum magnetoresistance since the separation between the impurities
and the 2D electrons suppresses the quantitative importance of $2k_F$
resistive scattering. Our maximum theoretical value of
$\rho(B_{\|})/\rho(0)$ for $z_i = 100 \AA$ of about 1.7 is in
excellent agreement with the corresponding experimentally observed
\cite{eight} value, lending support to our theoretical
screening picture.

In Fig. 4, which is perhaps the most important result presented in
this work, we show our calculated magnetoresistance as a function of
$B_{\|}$ (at a fixed 2D density of $10^{11}cm^{-2}$) for several
different temperatures -- the inset of Fig. 4 shows $\rho(T)$ at
$B_{\|} =0$ and 9T ($>B_c \approx 5T$ at $10^{11}cm^{-2}$ density).
The theoretical results in Fig. 4 are qualitatively similar to the
corresponding experimental results. The most important feature of
Fig. 4 is the metallic temperature dependence of $\rho(T;B_{\|})$ at
$B_{\|} >B_c$, which, although suppressed from the strong zero-field
metallic temperature dependence $\rho(T;B_{\|}=0)$ due to the
spin-polarization induced suppression of screening as the spin
degeneracy changes from the paramagnetic value of 2 for $B_{\|}=0$ to
the ``ferromagnetic'' value of 1 for $B_{\|} \ge B_c$, is still
strongly metallic in character with $\rho(T)$ changing by almost 50 \%
at $B_{\|} = 9T \approx 2B_c$. This is precisely the experimental
observation. In fact, our calculated temperature dependence of
$\rho(T;B_{\|}>B_c)$ is in excellent qualitative
agreement with the experimental results \cite{seven}.

The significance of our theoretical agreement with the observed
``metallic'' (i.e. $d\rho/dT >0$) resistivity for $B_{\|}>B_c$ arises
from the very specific and contrasting prediction for
$\rho(T;B_{\|}>B_c)$ in the so-called ``interaction theory''
\cite{twelve}, which purports to extend the screening theory by
calculating the electron-electron interaction effects 
(to leading order in
$T/T_F$) on the 2D transport properties to {\it all order}
(perturbatively) in interaction for an unrealistic zero-range
white-noise model of bare impurity disorder. The categorical
prediction \cite{twelve} of the interaction theory is that for a fully
spin polarized 2D system (i.e. for $B_{\|} \ge B_c$) 
the 2D resistivity must necessarily
manifest an ``insulating'' (i.e., $d\rho/dT <0$ for $B_{\|}>B_c$)
behavior. This
interaction theory prediction obviously disagrees qualitatively with
our results in Fig. 4, where $d\rho/dT >0$ for $B_{\|} >B_c$, as well
as with experimental observations \cite{seven} in the Si/SiGe based
high-mobility 2D electron systems. This qualitative disagreement
(agreement) between interaction (screening) theory and experiment
shows that the interaction theory is invalid in the 2D Si/SiGe systems
due to the long-range nature of the underlying impurity disorder.
We have also calculated (not shown) $\rho(B_{\|})$ as a function of
$B_{\|}$ at various densities, finding that $\rho(B_{\|})/\rho(0)$ at
various densities approximately scale with $B_{\|}/B_c$ as reported
experimentally \cite{eight}. But we find that this scaling is
non-universal and somewhat temperature dependent.

\begin{figure}
\epsfysize=2.3in
\centerline{\epsffile{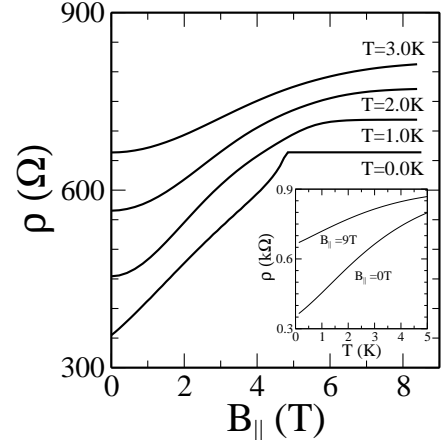}}
\vspace{0.5cm}
\caption{
Calculated magnetoresistance as a function of in-plane magnetic field
for different temperatures at a fixed carrier
density $n=10^{11} cm^{-2}$. Inset shows the resistivities
$\rho(T,B_{\|})$ at $B_{\|}=0$ and 9T as a function of temperature.
}
\label{fig4}
\end{figure}

In summary, we have developed a realistic screening theory for the
density, temperature, and parallel magnetic field dependence of
low-temperature 2D resistivity $\rho(T;n;B_{\|})$ in Si/SiGe electron
systems. Our results are in comprehensive qualitative
agreement (in $T$, $n$, and $B_{\|}$ dependence) with recent
experimental observations \cite{five,six,seven,eight}. In particular,
we get agreement with the experimental observation of a ``metallic''
(i.e., $d\rho/dT >0$) 2D resistivity in the completely spin-polarized
($B_{\|}>B_c$) system, which disagrees qualitatively with the opposite
prediction (i.e. $d\rho/dT <0$ for $B_{\|} >B_c$) of the interaction
theory \cite{twelve}. 
The agreement between our theory and experiments directly demonstrates
the key role of long-range Coulombic disorder in 2D metallicity.

This work is supported by ONR, NSF, and LPS.


\end{document}